# Decoupled magnetic and electrical switching in manganite trilayer

*L.E. Hueso \*, L. Granja [a]*

Department of Materials Science, University of Cambridge, Pembroke Street, Cambridge CB2 3QZ, UK

*P. Levy*

Departamento de Física, Comisión Nacional de Energía Atómica, Avenida Gral Paz 1499 (1650) San Martín, Buenos Aires, Argentina

*N.D. Mathur*

Department of Materials Science, University of Cambridge, Pembroke Street, Cambridge CB2 3QZ, UK

\* Corresponding author. E-mail: leh29@cam.ac.uk

[a] Permanent Address: Departamento de Física, Comisión Nacional de Energía Atómica, Avenida Gral Paz 1499 (1650) San Martín, Buenos Aires, Argentina

We report magnetic and electrical transport studies of an epitaxially grown trilayer thin film structure comprising $La_{0.59}Ca_{0.41}MnO_3$ sandwiched between $La_{0.67}Ca_{0.33}MnO_3$ electrodes. Since $La_{0.59}Ca_{0.41}MnO_3$ lies at the edge of the thin film ferromagnetic metallic phase field, phase separation effects are expected. These effects can explain the observed magnetic isotropy of the middle layer. By contrast, the electrode material is magnetically uniaxial. Easy axis magnetic field sweeps of the trilayer produce two sharp magnetic transitions, but only one sharp transition in current-in-plane resistance measurements.

The manganites, a class of mixed-valent manganese oxides, display a range of electronic and magnetic phases that can coexist even in a continuous crystal. This coexistence is caused by the strong interplay between the crystallographic, magnetic and electronic degrees of freedom.[1] $La_{1-x}Ca_xMnO_3$ (LCMO$x$%) is a typical manganite. The bulk phase diagram[2] is dominated by a ferromagnetic metallic phase in $x<0.5$, and a modulated phase[3,4] that was traditionally considered to be charge ordered in $x>0.5$.[5] The coexistence of these two phases has been observed through measurements of electrical resistivity and magnetization,[6,7] neutron diffraction,[8] magnetic resonance,[9,10] and electron microscopy[11] where the two phases are not always found to be mutually exclusive.[12]

Manganite thin film devices such as tunnel junctions[13,14] and exchange-biased multilayers[15,16] exploit the half-metallic character of certain compositions, but do not exploit phase separation. Here we attempt to exploit both phenomena by fabricating epitaxial trilayers in which the half metallic ferromagnet LCMO33 sandwiches LCMO41. This latter composition lies well within the ferromagnetic metallic phase in bulk samples[2]. However, epitaxial films of LCMO41 on $NdGaO_3$ (001) substrates (NGO) lie at the edge of the thin film ferromagnetic metallic phase field, and show strong evidence of phase separation with a ferromagnetic volume fraction of 45±5%.[17]

In this Letter, we demonstrate that this phase separated layer of LCMO41 is magnetically isotropic. Its presence in the trilayer magnetically decouples the two adjacent fully ferromagnetic LCMO33 layers. Normally, the independent switching of ferromagnetic domains leads to sharp jumps in magnetoresistance[18]. Our current-in-plane (CIP) magnetoresistance measurements reveal both sharp and diffuse jumps. We argue that the phase-separated nature of our interlayer decouples the magnetic and electrical switching fields in our device.

Using pulsed laser deposition, films of LCMO33 and LCMO41, and trilayer films of LCMO33/LCMO41/LCMO33 were grown on NGO substrates using conditions described elsewhere.[17] Three trilayers were grown in total with middle layer thicknesses of 10 nm, 20 nm and 40 nm. In each case the bottom and top LCMO layer thicknesses were 50 and 20 nm respectively. NGO was chosen as a substrate due to its small (~0.1%) mismatch with LCMO, and because it simplifies the micromagnetics by promoting uniaxial magnetic anisotropy.[19] However, the penalty for using this substrate is that magnetic measurements are difficult due to the strong paramagnetic signal arising from the neodymium ($J = 9/2$) ions.

Samples were characterized by x-ray diffraction (XRD) and atomic force microscopy (AFM). For all samples, ω-2θ scans around the NGO (004) peak show a Laue modulation indicating coherent growth (Fig. 1). The periodicity of the fringes matches the total nominal film thickness to within 2-3 nm. Moreover, reciprocal space maps around (116) pseudocubic reflections show that both in-plane film lattice parameters match the substrate. AFM measurements evidence step-terrace growth with unit cell terraces heights.

Vibrating sample magnetometer measurements at 50 K of a 20 nm LCMO41 film reveal both a reduced saturation magnetization and magnetic isotropy (Fig. 2). The observed saturation magnetization of 1.7±0.2 $\mu_B$/Mn represents 47±5% of the theoretical limit of 3.59 $\mu_B$/Mn, qualitatively supporting the previously proposed argument that a ferromagnetic phase coexists with a non-ferromagnetic phase leading to a metal-insulator transition at 150 K.[17] The observed magnetic isotropy can be explained if this ferromagnetic phase is distributed to form small (< 20 nm) islands[20] in which shape effects dominate the underlying magnetocrystalline anisotropy[19]. However, the observed metal-insulator transition[17] suggests the presence of ferromagnetic metallic percolation paths between these islands. (This conclusion is independently supported by the calculation that we present later.) We stress that the magnetically isotropic behavior of epitaxial LCMO41/NGO sharply contrasts the uniaxial anisotropy seen in the strongly ferromagnetic phase of epitaxial LCMO33/NGO, where the orthorhombic [100] corresponds to the easy axis.[19]

In Fig. 3 we present magnetic and electrical measurements taken at 50 K for the trilayer with middle layer thickness 20 nm. (The trilayers with middle layer thicknesses 10 nm and 40 nm behave similarly and will therefore not be discussed further.) In Fig. 3a, magnetic measurements along the orthorhombic [100] in-plane direction of the trilayer indicate the presence of three components in the hysteresis loop. The LCMO33 components show the expected easy axis behavior, and the magnitude of the associated sharp switching events permits coercive fields to be assigned to the 50 nm bottom (4.5±0.4 mT) and 20 nm top (10.7±0.5 mT) layers. The behavior of the LCMO41 component is equivalent to the isotropic measurements of the LCMO41 film (Fig. 2).

The sharp switching seen in Fig. 3a represents single domain switching between parallel and anti-parallel states of the top and bottom layers. The simple addition of the three hysteresis loops implies that all three layers are magnetically independent. On a major loop, before switching to the antiparallel state, the trilayer is essentially saturated. After switching, only the bottom layer is reversed. As the middle layer does not follow the magnetic behaviour of the LCMO33, we suggest that magnetic domain walls form between the bottom layer and the ferromagnetic regions of the phase separated middle layer. We therefore turn to electrical transport measurements, which have the potential to detect the presence of interfacial structures that cannot be resolved by magnetometry.

Current-in-plane (CIP) measurements of the trilayer were performed in the standard 4-terminal geometry along the orthorhombic [100] direction by depositing gold contact pads with a separation between the voltage pads of 2 mm. Fig. 3b demonstrates hysteresis in the magnetoresistance of our LCMO33/LCMO41/LCMO33/NGO trilayer. By contrast, no hysteresis is seen in similar measurements of LCMO33/NGO and LCMO41/NGO films. The most striking feature in Fig. 3b is that whilst the high resistance state arises from magnetic reversals of the bottom electrode (Fig. 3a), the low resistance state (associated with the high-field reversible regime) is not recovered when the magnetization of the top electrode reverses to yield the parallel state. Instead, the transition from high to low resistance mimics the magnetic changes in the middle layer (Fig. 3a).

The behavior seen in Fig. 3b may be understood qualitatively as follows. Switching to the antiparallel state of the top and bottom layers leads to the formation of domain walls at distributed locations near the lower interface, as discussed earlier. Domain walls in the manganites could possess a high electrical resistance due to mesoscopic phase separation.[21] These domain walls electrically isolate the bottom layer and produce a sharp increase in resistance. The subsequent magnetization reversal in the top layer brings the top and bottom layers into the parallel state around 10 mT, but at such low fields the middle layer remains essentially antiparallel to both. Therefore domain walls, which were previously only associated with the bottom interface, are now expected to also be present at the upper interface. However, the bottom layer remains electrically isolated and so no sharp switch to the low resistance state is observed. Instead, the low resistance state is approached gradually as the ferromagnetic regions in the middle layer reverse, removing domain walls near both interfaces and thus reintroducing current paths through the bottom layer as the magnetic field is increased.

We now present a quantitative analysis of the behavior seen in Fig. 3b. We approximate the resistivities of the component layers at 50 K to be isotropic and equal to the in-plane orthorhombic [100] values for LCMO33/NGO (1 μΩ.m) and LCMO41/NGO (100 μΩ.m).[17] Given the high value of the latter, we assume that current only flows in the LCMO41 layer through percolating pathways whose resistivity is equal to the resistivity of the LCMO33 film. We also assume that these pathways are lumped into just two regions, each centered under a voltage terminal. The domain walls form at the interfaces of these pathways with the LCMO33 layers, as described above. Assuming the 77 K domain wall resistance-area product ($\Delta RA \approx 10^{-13}$ $\Omega.m^2$) for LCMO30[21] we find that each of the two lumped pathways has an area of just ~1 $\mu m^2$. This is six orders of magnitude smaller than the active area of our device (2×2 $mm^2$), suggesting that CIP transport in the phase separated LCMO41 layer is mediated by very narrow current pathways between the postulated ferromagnetic islands.

In conclusion, epitaxy is preserved in all-manganite trilayers. The middle LCMO41 layer is magnetically isotropic despite the underlying magnetocrystalline anisotropy. We attribute this discrepancy to phase separation, and the ferromagnetic volume fraction is 47%. We argue that percolating pathways of fractional area $10^{-6}$ form between ferromagnetic islands. These pathways permit the current to flow via the bottom layer. The trilayer structure represents a device in which the magnetic and electrical switching are decoupled and may thus be independently controlled.

The authors thank J. Rivas and P.B. Littlewood for helpful discussions. This work was funded by the UK EPSRC, an EU Marie Curie Fellowship (LEH), Alβan Programme (LG) and The Royal Society.

**Figure Captions**

**Fig. 1** ω-2θ scans around the NGO (004) substrate peak for (bottom) a film of LCMO41/NGO and (top) a trilayer of LCMO33(20 nm)/LCMO41(20 nm)/LCMO33(50 nm)/NGO.

**Fig. 2** Magnetic hysteresis loops, corrected for the paramagnetic substrate of volume 7.5 mm$^3$, measured at 50 K along the three perpendicular <100> orthorhombic directions of a 20 nm LCMO41 film. The similarity between these three loops (and others taken in intermediate orientations) demonstrates the absence of magnetic anisotropy.

**Fig. 3** Magnetic and electrical measurements versus applied magnetic field for the LCMO33(20 nm)/LCMO41(20 nm)/LCMO33(50 nm)/NGO trilayer at 50 K. (a) Magnetization, corrected for the paramagnetic substrate of volume 7.5 mm$^3$, for the in-plane orthorhombic [100] direction. Easy axis switching of the top and bottom layers appears to be independent of the isotropic middle layer. (b) Current-in-plane resistance obtained at a measurement current of 10 μA. At all measurement fields, I-V curves were linear up to this current. The sharp switching corresponds to magnetic reversals of the bottom LCMO33 layer. No corresponding switch is seen for reversals of the top LCMO33 layer. The hysterestic regimes in (a) and (b) persist out to the same applied field.

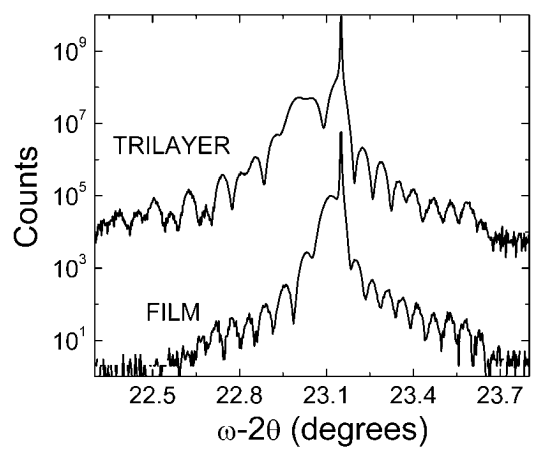

**Hueso** *et al.*

**Figure 1**

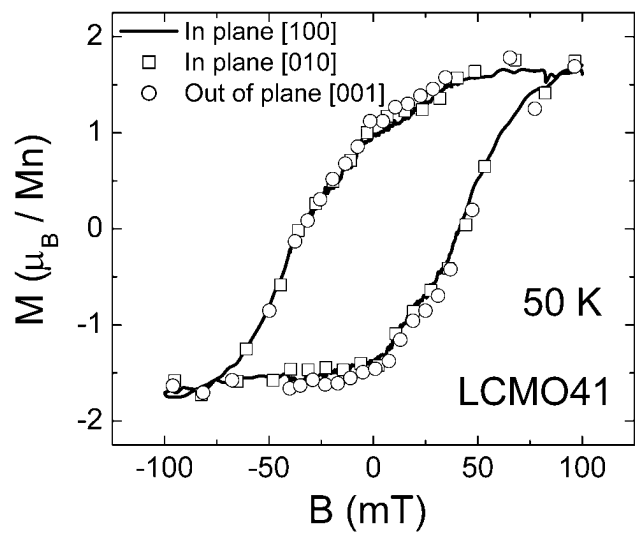

Hueso *et al.*

**Figure 2**

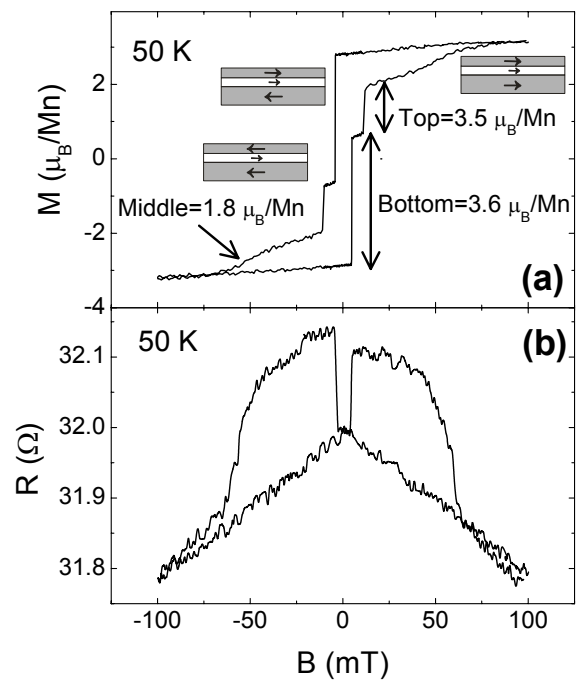

**Hueso** *et al.*

**Figure 3**